\documentclass{xpkas}

\def\year{2015} % publication year
\def\month{January} % publication month
\def\volume{00} % journal volume
\def\issue{0} % journal issue
\def\beginpage{1} % first page of article
\def\endpage{99} % last page of article
\def\received{October 31, 2014} % date paper was received by PKAS
\def\accepted{November 30, 2014} % date of acceptance
\date{Received \received ; accepted \accepted}

\usepackage{flushend} %% balance columns on last page
\newcommand\ion[2]{{#1}\,{\sc #2}} % ions: \ion{C}{iv} = C IV

\title{
Constraining supernova progenitors: an integral field spectroscopic survey of the explosion sites%\thanks{This document deals with the technical aspects of writing a PKAS paper, \emph{not} with editorial policy or scientific questions.}
}

\author[1,2]{H.~Kuncarayakti}
\author[3]{G~Aldering}
\author[4]{J.~P.~Anderson}
\author[5]{N.~Arimoto}
\author[6]{M.~Doi}
\author[1,2]{L.~Galbany}
\author[2,1]{M.~Hamuy}
\author[6]{Y.~Hashiba}
\author[4]{T.~Kruehler}
\author[7,8]{K.~Maeda}
\author[6]{T.~Morokuma}
\author[9]{T.~Usuda}
\affil[1]{Millennium Institute of Astrophysics, Casilla 36-D, Santiago, Chile}
\affil[2]{Departamento de Astronom\'ia, Universidad de Chile, Casilla 36-D, Santiago, Chile}
\affil[3]{Physics Division, Lawrence Berkeley National Laboratory, 1 Cyclotron Road, Berkeley, CA 94720, USA}
\affil[4]{European Southern Observatory, Alonso de Cordova 3107, Vitacura, Santiago, Chile}
\affil[5]{Subaru Telescope, National Astronomical Observatory of Japan, 650 North A'ohoku Place, Hilo, HI 96720, USA}
\affil[6]{Institute of Astronomy, Graduate School of Science, The University of Tokyo, 2-21-1 Osawa, Mitaka, Tokyo 181-0015, Japan}
\affil[7]{Department of Astronomy, Kyoto University, Kitashirakawa-Oiwake-cho, Sakyo-ku, Kyoto 606-8502, Japan}
\affil[8]{Kavli Institute for the Physics and Mathematics of the Universe (WPI), The University of Tokyo, Kashiwa, Chiba 277-8583, Japan}
\affil[9]{National Astronomical Observatory of Japan, Mitaka, Tokyo 181-8588, Japan}
\def\corrauthor{
\\
Hanindyo Kuncarayakti; \email{hanin@das.uchile.cl}
}

\def\runningauthor{
H. Kuncarayakti et al.
}

\def\runningtitle{
IFU survey of supernova explosion sites
}

\def\keywords{
supernovae, massive stars, stellar populations, integral field spectroscopy
}

\def\abstracttext{
We describe a survey of nearby core-collapse supernova (SN) explosion sites using integral field spectroscopy (IFS) technique, which is an extension of the work described in Kuncarayakti et al. (2013, AJ, 146, 30/31) . The project aims to constrain the SN progenitor properties based on the study of the SN immediate environment. The stellar populations present at the SN explosion sites are studied by means of integral field spectroscopy, which enables the acquisition of both spatial and spectral information of the object simultaneously. The spectrum of the SN parent stellar population gives the estimate of its age and metallicity. With this information, the initial mass and metallicity of the once coeval SN progenitor star are derived. While the survey is mostly done in optical, additionally the utilization of near-infrared integral field spectroscopy assisted with adaptive optics (AO) enables us to examine the explosion sites in high spatial details, down to a few parsecs. This work is being carried out using multiple 2--8 m class telescopes equipped with integral field spectrographs in Chile and Hawaii.
}

\begin{document}
\pkashead %% set title, authors, abstract, etc.

%%%%%%%%%%%%%%%%%%%%%%%%%%%%%%%%%%%%%%%%%%%%%%%%%%%%%%%%%%%%%%%%%%%%%
%%% BEGIN MAIN TEXT HERE %%%%%%%%%%%%%%%%%%%%%%%%%%%%%%%%%%%%%%%%%%%%
%%%%%%%%%%%%%%%%%%%%%%%%%%%%%%%%%%%%%%%%%%%%%%%%%%%%%%%%%%%%%%%%%%%%%

\section{Introduction\label{sec:intro}}
Despite their importance and large number of observed events, the progenitors of supernovae (SNe) are still not very clearly understood. There have been extensive efforts on both theoretical and observational grounds to constrain the physical properties and evolutionary status of the SN progenitors. Theoretical works \citep[e.g.][]{heger03,groh13} predict the evolution of massive stars and their final outcomes as different types of SNe, but these models still need to be verified with observational data and the fundamental question still lingers: \emph{is there any specific relation between SN type and the characteristics of the progenitor star?} 

During the last two decades, a handful of SN progenitors have been directly detected in pre-explosion archival Hubble Space Telescope images \citep[see][for a review]{smartt09araa}, but these detections are still rare and most SNe do not have pre-explosion data available. 
Thus far, SN Ib/c progenitors, which are believed to be either Wolf-Rayet stars stripped of their hydrogen envelope via metallicity-driven stellar wind or interacting massive binaries, have eluded discovery \citep[][however note the possible detection of the progenitor of the type-Ib SN iPTF13bvn, \citealt{cao13}]{eldridge13}. While powerful, archival direct detections are very rare and prone to the uncertainty on mass loss and the final stages of massive star evolution \citep{yoon12} as well as other complications such as extinction and metallicity assumption. As it is very difficult to improve the statistics due to the limited availability of usable pre-explosion images, there is necessity for alternative methods to constrain the progenitor properties better such as SN environment studies \citep[e.g.][]{anderson12,galbany14}. Thus, we initiated a statistical study of the SN parent stellar populations to characterize the SN progenitor stars. The initial results have been published in \citet{hk13a,hk13b} and in this paper we describe the efforts currently underway to improve the study and some examples of new results. The full account of the study with detailed results and thorough analysis will be published elsewhere (Kuncarayakti et al., in prep.).

\section{The sample}
Extending our works in \citet{hk13a,hk13b}, we employ the same strategy for the current investigation. We aim to observe the locations of nearby SNe ($\lesssim$20 Mpc) with integral field spectroscopy, in order to detect the parent stellar population of the SN progenitor and derive age and metallicity constraints from the extracted spectrum.

The initial sample of \citet{hk13a,hk13b} consists mainly of targets accessible from the northern hemisphere, and our currently ongoing observing campaign focuses on targets observable from the southern hemisphere ($\delta<30^\circ$). As for the initial sample, we used the Asiago Supernova Catalog \citep{barbon99} to select our observational targets. We selected core-collapse (non-Ia) SNe residing in host galaxies within $cz < 1500$ km s$^{-1}$ ($\lesssim20$ Mpc), having low host inclination \citep[$<65^\circ$, following][]{crowther13}, and not too old to have reasonable astrometric accuracy ($>1970$). SNe younger than 2013 were not included, in order to avoid contamination in the stellar population spectra from SN late-time emission. With these criteria, our survey is volume-limited and 66 SNe were selected for new observations. As of mid-2014, around half of the targets have been observed.

\begin{figure}[h]
\centering
\includegraphics[width=75mm]{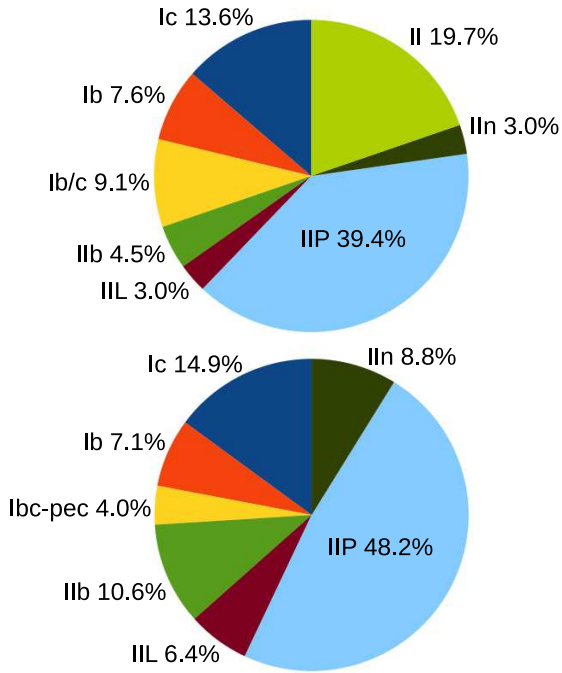}
\caption{Fractions of SN types in our new sample (\textit{upper panel}), compared to the the observed SN fractions from \citet[][\textit{bottom panel}]{smith11}. See text for details. \label{fig:frac}}
\end{figure} 
\vspace{5mm}

Figure \ref{fig:frac} shows the fractions of SN types in our new sample compared to the volume-limited observed fraction obtained from the Lick Observatory Supernova Search \citep[LOSS;][]{smith11}, which contains 80 core-collapse SNe within 60 Mpc. In our sample, there are SNe classified as types "II" or "Ib/c" in the Asiago Supernova Catalog, which are not present in the diagram of LOSS fractions. These type-II or Ib/c SNe were not very precisely typed, thus may actually consist of other more specific subclasses such as type-IIP/IIL/IIb/IIn or type-Ib/Ic. \citet{smith11} include peculiar broad-lined type Ib/c SNe, which are not present in our sample. Despite the differing classification details, it is evident that the fractions of SN types in our sample resembles those of \citet{smith11}. SNe IIP are the most frequent, comprising $\sim$40--50\% of core-collapse SNe, and hydrogen-poor SNe which encompass types-Ib, Ic, Ib/c, and IIb contribute to $\sim$35\% of the whole core-collapse SN population.

\section{Data collection}
As described in \citet{hk13a,hk13b}, the initial dataset was obtained using UHawaii2.2m/SNIFS \citep{aldering02,lantz04} and Gemini-N/GMOS \citep{allington02,hook04} atop Mauna Kea, Hawaii. The extension of the project currently uses telescopes stationed in Chile for the same purpose, but covering the southern targets. Multiple integral field spectrographs at VLT are being employed \citep[VIMOS, SINFONI, MUSE;][]{levefre03,eisenhauer03,bonnet04,bacon10}, as well as Gemini-S/GMOS and Magellan/IMACS \citep{bigelow98,schmoll04}. We use GMOS-S with identical instrument configuration as GMOS-N: 1-slit IFU mode with B600 grating. For IMACS we use IFU mode at f/2 (short camera) with 300 l/mm grism. GMOS and IMACS give a similar spatial sampling of 0.2" and field of view of 5"$\times$7.5". For VIMOS observations we use the MR grating IFU mode at 0.33" sampling. Table \ref{tab:ifu} lists the characteristics of the integral field spectrographs used in this study. The observations were started in April 2014 and consist of multiple observing runs throughout 2014 and 2015.

\begin{table}[t]
\caption{Instrument configurations used in this study. \label{tab:ifu}}
\centering
\begin{tabular}{lcccc}
\toprule
Instrument    & Spaxel & FoV & $\lambda$ range & R \\
\midrule
SNIFS & 0.43" & 6.4"$\times$6.4" & 330--930 nm & $\sim$1000 \\
VIMOS & 0.33" & 13"$\times$13" & 480--1000 nm & $\sim$1000 \\
MUSE & 0.2" & 60"$\times$60" & 480--930 nm & $\sim$2500 \\
GMOS-N/S & 0.2" & 5"$\times$7.5" & 400--680 nm & $\sim$1700 \\
IMACS & 0.2" & 5"$\times$7.5" & 400--900 nm & $\sim$1700 \\
SINFONI & 0.1" & 3"$\times$3" & 1.95--2.45 $\mu$ & $\sim$4000 \\
\bottomrule
\end{tabular}
\tabnote{
\textsc{Note:} SNIFS, VIMOS, and MUSE spaxels are square-shaped; the indicated spaxel sizes are the sides of the squares. GMOS and IMACS spaxels are haxagons; the indicated spaxel sizes are the diameters of the hexagon. SINFONI spaxels are rectangles with size 0.05"$\times$0.1".
}
\end{table}
\vspace{1cm}

While this study is mostly done in optical, additional observations in the infrared were also obtained using VLT/SINFONI. The incorporation of SINFONI in this study enables us to obtain high spatial resolution observations with the help of the adaptive optics system. We use the 100mas setting for SINFONI, which gives high spatial resolution of 0.05"$\times$0.1" per spaxel (not seeing-limited, which is the case for the optical observations) within its 3"$\times$3" field of view. This has the purpose of resolving the stellar populations better to map the star formation history of the SN explosion site. With better characterization of star formation history, the constraint on the stellar population age -- which has been usually assumed to be formed by an instantaneous burst -- could be improved. 

We further improve this study by adding MUSE observations. MUSE, a second-generation instrument at VLT, is capable of taking $\sim$ 90 000 spectra in one exposure with its wide IFU field of view of 1'$\times$1'. Using this large field of view, the whole host galaxies of the SN can also be observed with IFS, simultaneously with the SN explosion sites. Besides its wide field of view, MUSE also has high spatial resolution of 0.2"/spaxel, thus enabling us to resolve individual stellar populations within the galaxy with spatial resolution comparable to that of the other smaller-format IFU spectrographs used in this study. The parent stellar population of the SN will then be compared to other stellar populations present in the host galaxy. We will also map various parameters such as age and metallicity of stellar populations across the SN host galaxy, in fine spatial details. A separate pilot study using MUSE data taken during the instrument Science Verification phase is currently underway (Galbany et al., in prep.).

\section{Example results \& analysis}

The raw data collected from the IFS observations were recorded by the detector in the form of arrays of two-dimensional spectra. Subsequent reduction, extraction, and remapping steps were then performed using the available softwares of each respective instruments, in order to obtain the final product in the form of datacubes. These so-called IFS datacubes consist of information in both spatial (\textit{x,y}) and spectral ($\lambda$) dimensions. In the datacube, every spectra corresponding to each spatial pixels (spaxels) of the integral field unit were remapped onto the instrument field of view to reconstruct the object appearance at the focal plane. Therefore it is possible to reproduce the image of the object in any wavelength bin along the $\lambda$ axis of the datacube, or extract the spectrum from any position within the field of view.

\begin{figure}[b]
\centering
\includegraphics[width=85mm]{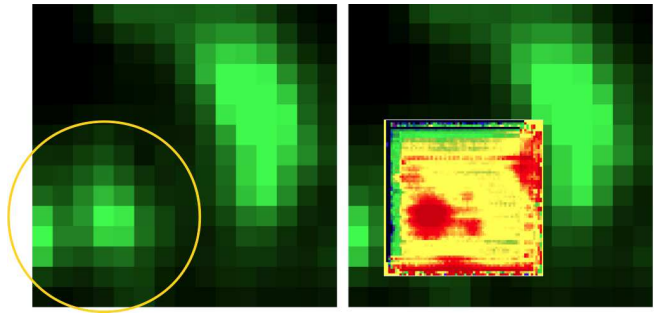}
\caption{(\textit{Left}) Integrated-light optical view of a SN site, taken with SNIFS, with the SN position indicated inside the 2"-radius yellow error circle. The image spans 220 pc in projected linear size along its sides (15 pc/spaxel scale). (\textit{Right}) The same image superposed with $K$-band image generated from SINFONI observation. The scale for the $K$-band image is 3.4 pc/spaxel in vertical direction, and 1.7 pc/spaxel in horizontal direction. \label{fig:sinfoex}}
\end{figure}

\begin{figure*}[]
\centering
\includegraphics[width=170mm]{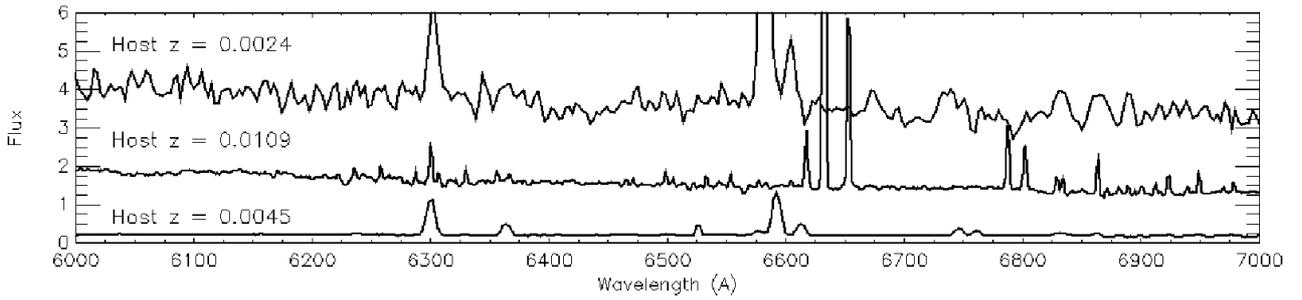}
\caption{An example of spectra around H$\alpha$ region extraced from one spaxel of (from top to bottom) SNIFS, MUSE, and VIMOS datacubes, in arbitrary flux values. Spectra are not background-subtracted; the strong [\ion{O}{I}]$\lambda$6300 emissions are night sky emission lines. \label{fig:spec}}
\vspace{5mm} %% add extra space ONLY when figures/tables are "colliding"!
\end{figure*}

Figure~\ref{fig:sinfoex} shows an example of image produced from a datacube. The image was generated by collapsing the datacube in the wavelength direction, thus represents an integrated-light image within the instrument's spectral response. Figure \ref{fig:sinfoex} also illustrates how the AO-assisted SINFONI observations improve the spatial resolution for the target. The stellar population associated with the SN (located within the SN position error circle) is now resolved into four individual sources. With this kind of data, it is possible to examine the age of the individual stellar populations and derive the star formation history of the SN explosion sites.

Each of the spaxels on the IFU image yield a spectrum, as shown in Figure~\ref{fig:spec}. As typical with young, star-forming regions, the spectra extracted from a SN explosion site usually show strong emission lines originating from ionized gas being excited by young, hot, massive stars. The metallicity is derived by employing the strong line method to the spectrum. Strong emission lines in the spectra such as H$\alpha$, [\ion{N}{II}]$\lambda$6584, H$\beta$, and [\ion{O}{III}]$\lambda$5007 are measured and used for calculating the N2 and O3N2 indices \citep{pp04}, which are used to estimate the metallicity. The age of the stellar population is estimated from the equivalent width of H$\alpha$ emission line, and in several cases of Ca-triplet absorption lines around $\lambda$8600. For the $K$-band infrared data, age is derived from the equivalent widths of Br$\gamma$ emission and CO 2.3$\mu$ absorption. These spectral age indicators are compared with simple stellar population (SSP) model Starburst99 \citep{leitherer99} to derive the stellar population age. With this method, the age and metallicity of the once coeval SN progenitors are derived from their parent stellar populations. As the lifetime of a star is governed mainly by its initial mass, the derived age can be converted into initial mass by comparing to stellar evolution models.

\section{Summary}

We briefly describe a survey of nearby SN sites using integral field spectroscopy technique, in order to constrain the physical properties of the SN progenitors such as initial mass and metallicity. This work is an extension of the study presented in \citet{hk13a,hk13b}. Sample selection, a part of the data collection as well as example results are presented. With this large collection of IFS data we aim to perform statistical analysis to the SN environment and progenitors, to provide better constraints and improve our understanding on different SN types and progenitor populations (Kuncarayakti et al., in prep.).

\acknowledgments

%We are grateful to all past and current PKAS authors for their trust and their support.
Support for HK, LG, and MH is provided by the Ministry of Economy, Development, and Tourism's Millennium Science Initiative through grant IC12009, awarded to The Millennium Institute of Astrophysics, MAS. HK acknowledges support by CONICYT through FONDECYT grant 3140563, and LG through grant 3140566. HK further acknowledges invitation and support from the conference organizers.

\end{document}